\documentclass[12pt,letterpaper]{article}
\usepackage{times}
\usepackage{amssymb}
\usepackage{t1enc}
\usepackage[latin1]{inputenc}
\usepackage[english]{babel}
\usepackage{graphicx}
\renewcommand{\baselinestretch}{1.75}

\begin{document}

\title{Test of Bivariate Independence Based on Angular Probability Integral Transform with Emphasis on Circular-Circular and Circular-Linear Data}
\author{Fern\'andez-Dur\'an, J.J. and Gregorio-Dom\'inguez, M.M. \\
ITAM\\
E-mail: jfdez@itam.mx}
\renewcommand{\baselinestretch}{1.00}
\date{}
\maketitle

\renewcommand{\baselinestretch}{1.75}

\begin{abstract}
The probability integral transform (PIT) of a continuous random variable $X$ with distribution function $F_X$ is a uniformly distributed random variable $U=F_X(X)$.
We define the angular probability integral transform (APIT) as $\theta_U = 2 \pi U = 2 \pi F_{X}(X)$, which corresponds to a uniformly distributed angle on the unit
circle.
For circular (angular) random variables, the sum of absolutely continuous independent circular uniform random variables is a circular uniform random variable, that is, the
circular uniform distribution is closed under summation, and it is a stable continuous distribution on the unit circle. If we consider the sum (difference) of the
angular probability integral transforms of two random variables, $X_1$ and $X_2$, and test for the circular uniformity of their sum (difference), this is equivalent to the test of independence of the original variables. In this study, we used a flexible family of nonnegative trigonometric sums (NNTS) circular distributions, which include
the uniform circular distribution as a member of the family, to evaluate the power of the proposed independence test; we complete this evaluation by generating samples from NNTS alternative
distributions that may be at a closer proximity with respect to the circular uniform null distribution.
\end{abstract}


\newpage
%\textbf{Keywords}: .

%\renewcommand{\baselinestretch}{1.75}

\section{Introduction}

Testing for independence is one of the most important tasks in statistics, for example, when constructing the joint distribution of a set of random variables or considering the conditional dependence of one variable in terms of other variables as in regression models.

According to Herwatz and Maxand (2020), one can consider the following tests of independence: bivariate (pairwise), mutual, and groupwise independence tests. Hereinafter, we consider only absolutely continuous random variables. A random variable with a density function with support on an interval of
the real line is a linear random variable, and a random variable with support on the unit circle is a circular random variable. 
The distribution function of a circular random variable is a periodic function and then, it has an arbitrary starting direction. In our case, we considered that all the circular random variables are defined on the interval $(0,2\pi]$ with a common starting direction and are angles in radians measured all counterclockwise or clockwise. In this sense, if $\Theta$ is a circular random variable with starting direction zero and measured in radians counterclockwise, 
\[
F_{\Theta}(\theta) = P(0 < \Theta \le \theta) \mbox{ for } 0 \le \theta \le 2\pi
\]
and
\[
F_{\Theta}(\theta+2k\pi)=F_{\Theta}(\theta)
\]
for $k$ an integer number. Thus, the density function for an absolutely continuous circular random variable is a periodic function such that $f_{\Theta}(\theta) \ge 0$, $f_{\Theta}(\theta+2k\pi)=f_{\Theta}(\theta)$ and
$\int_{0}^{2\pi}f_{\Theta}(\theta)d\theta=1$. For two random variables, $X_1$ and $X_2$, bivariate
(pairwise) independence tests have a null hypothesis, $H_0: F_{X_1,X_2}(x_1,x_2)=F_{X_1}(x_1)F_{X_2}(x_2)$, where $F_{X_1,X_2}(x_1,x_2)=P\{X_1 \le x_1, X_2 \le x_2\}$ is the
bivariate joint distribution function and $F_{X_1}(x_1)=P\{X_1 \le x_1\}$ and $F_{X_2}(x_2)=P\{X_2 \le x_2\}$ are the corresponding marginal distribution functions. A mutual independence test for a set of random variables, $X_1, X_2, \ldots, X_d$ considers the null hypothesis
$H_0: F_{X_1, X_2, \ldots, X_d}(x_1, x_2, \ldots, x_d) = \prod_{k=1}^{d}F_{X_k}(x_k)$ where $F_{X_1, X_2, \ldots, X_d}(x_1, x_2, \ldots, x_d)$ is the joint distribution and $F_{X_k}(x_k)$
for $k=1, 2, \ldots, d$ are the marginal univariate distribution functions. Finally, groupwise independence tests consider the independence between disjoint subsets of random variables. If the parametric functional forms of $F_{X_1, X_2, \ldots, X_d}$ and $F_{X_k}$ for $k=1,2, \ldots, d$ are specified, a likelihood-ratio test of independence from a sample of random vectors $\underline{X}_i=(X_{i1},X_{i2}, \ldots, X_{id})^{\top}$ of size $n$ ($i=1,2, \ldots, n$), can be constructed.
The most commonly used test for independence is the chi-squared test of
independence for contingency tables, which is not adequate when dealing with absolutely continuous random variables.
As an alternative to the likelihood ratio independence test, other tests for independence were developed by considering nonparametric (distribution free) methods, rank tests
(see Hoeffding, 1948 and Kendall and Stuart,1951), and measures of dependence (association) derived from the empirical copula process (Deheuvels, 1981; Genest and
R\'{e}millard, 2004; Genest and Verret, 2005; Genest et al., 2019; Roy, 2020 and Roy et al., 2020). The empirical copula, $C_n$, for a vector of $d$ absolutely continuous
linear random variables, $\underline{X}^\top=(X_1, X_2, \ldots, X_d)^\top$, and a sample of size $n$ is defined as follows:
\begin{equation}
C_n(u_1,u_2, \ldots, u_d)=\frac{1}{n} \sum_{j=1}^n I(\hat{F}_1(X_{i1}) \le u_1, \hat{F}_2(X_{i2}) \le u_2, \ldots, \hat{F}_d(X_{id}) \le u_2)
\end{equation}
where $I()$ is an indicator function, which is equal to one if the condition in its argument is satisfied and zero otherwise, and $\hat{F}_1, \hat{F}_2, \ldots, \hat{F}_d$ are the
empirical distribution functions of the random variables $X_1, X_2, \ldots, X_d$. A test of independence based on the distance between the empirical copula and
independence
copula for absolutely continuous linear random variables is implemented in the $R$ package $copula$ (Hofert et al., 2022 and Kojadinovic and Yan, 2010).

Among the nonparametric tests of independence, there is a family of tests based on some function of the empirical independence process, which is defined as the distance between
the empirical joint distribution function and the product of the empirical univariate distribution functions. Historically, the most commonly used functionals have been the
Cram\'{e}r-von Mises and Kolmogorov-Smirnov functionals (refer to Blum, Keifer and Rosenblatt, 1961; DeWet, 1980 and Deheuvels, 1981).
For example, Hoeffding (1948) considered the Cram\'{e}r-von Mises functional to generate a rank test of independence between two random variables. Modern rank tests of
independence have been developed by Kallenberg and Ledwina (1999). Kernel-based methods have also been used to estimate the empirical independence process, as in Pfister et al.
(2018). Mardia and Kent (1991) used the general Rao score test to generate
independence tests. Cs\"{o}rg\H{o} (1985) developed independence tests based on the
multivariate empirical characteristic function, and Einmahl and McKeague (2003) developed the tests based on empirical likelihood. The measures of dependence derived from entropy were defined
by Joe (1990) and from mutual information by Berrett and Samworth (2019).

When constructing tests of independence, one can take advantage of the characteristics of the multivariate joint distribution. For example, for the multivariate normal distribution, one can test for independence by testing for an identity correlation matrix. Some of these pairwise tests are the Pearson (1920) product moment correlation coefficient test, Kendall (1938) rank correlation coefficient test, and Spearman (1904) rank correlation coefficient test. Of course, for a pair of Gaussian random variables, rejecting null correlation implies rejecting pairwise
independence, but
applying pairwise independence (correlation) tests is not adequate to test for mutual independence for a set with more than two Gaussian random variables. The Wilks test (Wilks,
1935) is an optimal test of independence for multivariate Gaussian populations, and for the case of a bivariate groupwise independence test for the vectors
$\underline{X}_{D_1}$ and $\underline{X}_{D_2}$ with $\underline{X}=\underline{X}_{D_1 \bigcup D_2}^\top=(\underline{X}_{D_1},\underline{X}_{D_2})^\top$, this test considers the
following test statistic for a sample of size $n$,
\begin{equation}
W = \frac{|\hat{\Sigma}_{D_1 \bigcup D_2}|}{|\hat{\Sigma}_{D_1}||\hat{\Sigma}_{D_2}|}
\end{equation}
where $\hat{\Sigma}_{D_1 \bigcup D_2}=\sum_{j=1}^{n}(\underline{x}_j - \bar{\underline{x}})(\underline{x}_j - \bar{\underline{x}})^\top$ is the estimated covariance matrix
of the complete vector of observations $\underline{x}=\underline{x}_{D_1 \bigcup D_2}$ which is partitioned into $\hat{\Sigma}_{D_1}$ and $\hat{\Sigma}_{D_2}$
with $\hat{\Sigma}_{D_1}=\sum_{j=1}^{n}(\underline{x}_{D_1,j} - \bar{\underline{x}}_{D_1})(\underline{x}_{D_1,j} - \bar{\underline{x}}_{D_1})^\top$ and
$\hat{\Sigma}_{D_2}=\sum_{j=1}^{n}(\underline{x}_{D_2,j} - \bar{\underline{x}}_{D_2})(\underline{x}_{D_2,j} - \bar{\underline{x}}_{D_2})^\top$. The statistic $W$ then
measures the extent of the distance between the determinant of $\hat{\Sigma}_{D_1 \bigcup D_2}$ and the product of the determinants of $\hat{\Sigma}_{D_2}$ and $\hat{\Sigma}_{D_2}$. The equality
relationship is satisfied in the multivariate Gaussian population under the null hypothesis of independence between $\underline{X}_{D_1}$ and $\underline{X}_{D_2}$.
Asymptotically and under regularity conditions, $-n\ln(W)$ follows a chi-squared distribution with $\sharp D_1\sharp D_2$ degrees of freedom, where $\sharp D_1$ and $\sharp
D_2$ are the cardinalities of sets $D_1$ and $D_2$, respectively.

For the circular-circular (angular-angular) and circular-linear (angular-linear) cases, in which the objective is to test for bivariate independence
between two circular random variables and one circular and one linear random variable, respectively, independence tests have been developed by considering
the specification of measures of dependence and by
studying their (asymptotic) distributions. By applying Kendall's tau and Spearman's rho general measures
of dependence based on the concept of concordance, or to the construction of distribution-free correlation coefficients based on ranks to a pair of circular random variables
or a circular and a linear random variables, tests of independence were developed by Fisher and Lee (1981, 1982 and 1983) and revised by Fisher (1993)
and Mardia and Jupp (2000).

The objective of this study is to develop a test of bivariate (pairwise) independence for two random variables by considering the angular probability transform of each variable, which corresponds to circular uniform distributions on $(0,2\pi]$, and by considering an additional result of the theory of circular statistics (see Fisher, 1993;
Upton and Fingleton, 1989; Mardia and Jupp, 2000 and Jammalamadaka and SenGupta, 2001). The test was evaluated by using flexible nonnegative trigonometric sums (NNTS)
distributions (Fern\'{a}ndez-Dur\'{a}n, 2004b, 2007). Although the proposed test of independence is a general one, it is especially suitable to test for independence
in the circular-circular and circular-linear cases. Thus, a measure of dependence was developed.

This paper is divided into six sections, including the introduction. In the second section, the Johnson and Wehrly (1977) model is presented as a motivation for performing the test of bivariate independence for two random variables, and here, the theory of NNTS circular distributions is included. The third section presents the proposed bivariate
independence test, a measure of dependence, and its application to simulated data to study the power of the test. In this section, we explain how to extend
the test of bivariate independence to a test of groupwise independence for two disjoint subsets of random variables. The fourth section includes a simulation study to
evaluate the power of the proposed test in the linear-linear, circular-linear, and circular-circular cases. The fifth section describes the application of the proposed
independence test to real datasets. Finally, conclusions are presented in the sixth section.

\section{Bivariate Johnson and Wehrly Model and NNTS Family of Circular Densities}

Sklar's (1959) theorem specifies that the joint cumulative distribution function of two continuous random variables, $X_1$ and $X_2$, with corresponding cumulative distribution functions $F_{X_1}(x_1)$ and $F_{X_2}(x_2)$ can be expressed as
\begin{equation}
F_{X_1,X_2}(x_1,x_2)=C(F_{X_1}(x_1),F_{X_2}(x_2))
\label{copuladist}
\end{equation}
where $C(u,v)$ is the copula (distribution) function that corresponds to a bivariate joint distribution of two identically distributed uniform random variables on the interval $(0,1]$, $U=F_{X_1}(X_1)$ and $V=F_{X_2}(X_2)$. A linear-linear copula function must be an increasing function satisfying $C(u,1)=u$, $C(1,v)=v$, $C(0,v)=C(u,0)=0$ (see
Nelsen, 1999). By differentiation of Equation \ref{copuladist}, one obtains the joint density function of $X_1$ and $X_2$, $f_{X_1,X_2}(x_1,x_2)$, as
\begin{equation}
f_{X_1,X_2}(x_1,x_2)=c(F_{X_1}(x_1),F_{X_2}(x_2))f_{X_1}(x_1)f_{X_2}(x_2)
\label{copuladensity}
\end{equation}
where $f_{X_1}(x_1)$ and $f_{X_2}(x_2)$ are the marginal density functions of $X_1$ and $X_2$, respectively. The function $c(u,v)$ is the copula density function defined as
$c(u,v)=\frac{\partial^2 C(u,v)}{\partial u \partial v}$. Johnson and Wehrly (1977) and Wehrly and Johnson (1980) proposed a large family of joint density functions for circular-circular 
($\Theta_1$ and $\Theta_2$) and circular-linear ($\Theta$ and $T$) random vectors in the following way:
\begin{equation}
f_{\Theta_1,\Theta_2}(\theta_1,\theta_2)= 2\pi g(2\pi(F_{\Theta_1}(\theta_1) \pm F_{\Theta_2}(\theta_2))f_{\Theta_1}(\theta_1)f_{\Theta_2}(\theta_2)
\label{jwcircularcircular}
\end{equation}
and
\begin{equation}
f_{\Theta,T}(\theta,t)= 2\pi g(2\pi(F_{\Theta}(\theta) \pm F_{T}(t))f_{\Theta}(\theta)f_{T}(t).
\label{jwcircularlinear}
\end{equation}
Fern\'{a}ndez-Dur\'{a}n (2004) identified the structure of the Johnson and Wehrly model in terms of the theory of copula functions through Sklar's (1959) theorem (see
Nelsen, 1999) satisfying
\begin{equation}
c(u,v)=2\pi g(2 \pi (u \pm v)).
\end{equation}
The function $g$ must be the density function of an angular (circular) random variable on the interval $(0,2 \pi]$. For the case of circular-circular and circular-linear bivariate copulas, the
function $c$ must also be periodic in their circular arguments to satisfy the periodicity of the density function of a circular random variable, which is the reason that in the Johnson and Wehrly model the joining function $g$ is the density of a circular random variable.

Johnson and Wehrly derived bivariate circular-circular and circular-linear models by considering conditional arguments. When function $g$
corresponds to a uniform circular density on the circle, $g(\theta)=\frac{1}{2 \pi}$, the joint density of the Johnson and Wehrly model corresponds
to the independence case in which the joint density of the circular-circular (circular-linear) model is the product of the marginal univariate densities.

This property of the Johnson and Wehrly model motivated our independence test by approximating the circular density function $g$ with
a density function from the NNTS family of densities (Fern\'{a}ndez-Dur\'{a}n, 2004b and 2007), which includes uniform circular density as a particular case. In addition, by using the NNTS family
of circular densities, it is possible to generate joining functions $g$ that are in closer proximity to the circular uniform density as desired; this is explained below. Pewsey and Kato (2016) developed a goodness-of-fit test for the circular-circular model of Wehrly and Johnson (1980) in Equation \ref{jwcircularcircular}. Their test considers the independence between each circular random variable, $\Theta_1$ and $\Theta_2$, and the argument of the joining function $g$, $\Omega=2 \pi \{F_{\Theta_2}(\theta_2) \pm F_{\Theta_1}(\theta_1)\}(mod 2\pi)$. Similar to this paper, their independence test is implemented by testing for the toroidal uniformity of the two bivariate random vectors, $(2\pi F_{\Theta_1}(\theta_1),2 \pi F_{\Omega}(\omega))^\top$ and $(2 \pi F_{\Theta_2}(\theta_2),2 \pi F_{\Omega}(\omega))^\top$.

The circular density function based on
nonnegative trigonometric sums (NNTS) for a circular (angular) random variable $\Theta \in (0,2\pi]$ (see Fern\'{a}ndez-Dur\'{a}n, 2004b) is defined as
\begin{equation}
f_{\Theta}(\theta ; M, \underline{c}) = \frac{1}{2\pi} \left|\left|\sum_{k=0}^{M}c_k e^{ik\theta}\right|\right|^2=
\frac{1}{2\pi} \sum_{k=0}^M\sum_{l=0}^M c_k\bar{c}_l e^{i(k-l)\theta}
\end{equation}
where $i=\sqrt{-1}$, $c_k$ are complex numbers $c_k=c_{rk} + ic_{ck}$ for $k=0, \ldots, M$ and $\bar{c}_k=c_{rk} - ic_{ck}$ is the conjugate of $c_k$.
To have a valid density function, $f_{\Theta}(\theta; M,\underline{c})$, that integrates to one,
\begin{equation}
\sum_{k=0}^M ||c_k||^2 = 1.
\label{cconst}
\end{equation}
where $||c_k||^2=c_{rk}^2 + c_{ck}^2$. The parameter space of the vector of parameters, $\underline{c}=(c_0,c_1, \ldots, c_M)^\top$, is a subset of the surface of a hypersphere since $\underline{c}$ and $-\underline{c}$ gives the same NNTS density and the conjugate of $\underline{c}$ written in reverse order also gives the same NNTS density as $\underline{c}$.
Then, for identifiability of the parameter vector $\underline{c}$ and given Equation \ref{cconst}, the following constraints are imposed:
$c_{c0}=0$ and $c_{r0} \ge 0$, i.e., $c_0$ is a nonnegative real number and $c_0^2 \ge ||c_M||^2$. 
There is a total of $2M$ free parameters $c$, and $M$ is an additional parameter that determines the total number of terms in the sum defining the density, and the parameter is
related to the maximum number of modes that the density can have. We note that an NNTS model with $M=M_1$ is nested on NNTS models with $M=M_2$ such that $M_2>M_1$. 
The circular uniform density on $(0,2\pi]$ corresponds to an NNTS density with $M=0$,
$f_{\Theta}(\theta;M=0,\underline{c})=\frac{1}{2\pi}$, and it is nested in all NNTS models with $M>0$. Equivalently, for an NNTS model with $M>0$, as $c_0$ approaches 1, the NNTS density converges to the circular uniform density; equivalently, for an NNTS density with $\underline{c}=(1,0,0,\ldots,0)^\top$, that is, the situation with $c_0=1$ and the other elements of
$\underline{c}$ equal to zero corresponds to the circular uniform density. This property is used to evaluate the power of the proposed independence test by generating samples from NNTS alternative densities with values of $c_0$ as close to one as desired. 
Fern\'{a}ndez-Dur\'{a}n and Gregorio-Dom\'{i}nguez (2010) developed an efficient algorithm based on optimization on manifolds to obtain the maximum likelihood
estimates of the $c$ parameters. This algorithm is included with other routines for the analysis of circular data based on NNTS models in the free $R$ (R Core Team, 2021) package \emph{CircNNTSR} (Fern\'{a}ndez-Dur\'{a}n and Gregorio-Dom\'{i}nguez, 2016).

\section{Proposed Test for Bivariate Independence}

For absolutely continuous independent and identically distributed (i.i.d.) circular uniform random variables, $U_1, U_2, \ldots, U_d \sim U(0,2 \pi]$, consider $\sum_{k=1}^{d} U_k \sim U(0,2 \pi]$, that is, the sum of i.i.d. circular uniform random variables is also circular and uniformly distributed. This is a consequence of the fact that the characteristic function of a circular uniform random variable, $\psi(t)=E(e^{it\Theta})$, is equal to one when $t=0$ and equal to zero elsewhere ($t \ne 0$). Then, if the sum of circular uniform random variables is not circular uniformly distributed, this implies that the circular uniform random variables are not independent (see Mardia and Jupp, 2000, p. 35).
The proposed test for independence is based on this result by considering the angular probability integral transform
of arbitrary (linear or circular) absolutely continuous random variables. Let $X_1, X_2, \ldots, X_d$ be $d$ arbitrary absolutely continuous random variables. The angular
probability transform of $X_k$ is defined as the angular (circular) random variable $APIT(X_k)=2 \pi F_{X_k}(X_k)$, which is uniformly distributed on the unit circle. By considering the null hypothesis of mutual joint independence, $APIT(X_1)$, $APIT(X_2)$, $\ldots, APIT(X_d)$ are $i.i.d.$ $U(0,2 \pi]$, then $\sum_{k=1}^{d}\pm APIT(X_k) \sim U(0,2
\pi]$ is also circular and uniformly distributed. The proposed test for bivariate independence is based on testing for the circular uniformity of $APIT(X_1)+APIT(X_2)$
($APIT(X_1)-APIT(X_2)$) for absolutely continuous (circular or linear) random variables $X_1$ and $X_2$. Testing for bivariate independence is equivalent to testing for
uniformity of the sum (difference) of the angular probability integral transforms. To test for circular uniformity of the
sum (difference) of the angular integral transforms, we considered the tests of Rayleigh and Pycke (Pycke, 2010). The Rayleigh test considers an alternative unimodal
circular density and has a test statistic for a sample $\theta_1, \theta_2, \ldots, \theta_n$, which is defined as follows (Mardia and Jupp, 2000): follows:
\begin{equation}
T_{RT}=2n\bar{R}^2
\end{equation}
where $\bar{R}$ is the sample mean resultant length. The statistic $T_{RT}$ asymptotically has a chi-squared distribution with two degrees of freedom. The p values of the Rayleigh test used in this paper correspond to the approximation given by Fisher (1993, p. 70) that includes a correction for small sample sizes.
The Pycke test considers an alternative multimodal density, and its test statistic for a sample $\theta_1, \theta_2, \ldots, \theta_n$ is defined as follows:
\begin{equation}
T_{PT}=\left(\frac{1}{n}\right)\sum_{i=1}^{n}\sum_{j=1}^{n} \left(\frac{2(\cos(\theta_i-\theta_j)-\sqrt{0.5})}{1.5 - (2\sqrt{0.5}\cos(\theta_i-\theta_j))}  \right).
\end{equation}
The critical values of the Pycke test are obtained via simulation.

The steps of the proposed independence test for two absolutely continuous random variables, $X_1$ and $X_2$ are as follows. First, the values of the
pseudo-observations (empirical distributions), $\hat{F}_{X_1}$ and $\hat{F}_{X_2}$ were calculated from the observed values of each random variable. Second, the angular
probability integral transforms (APITs) were calculated by multiplying the pseudo-observations by 2$\pi$, ($2\pi\hat{F}_{X_1}$ and $2\pi\hat{F}_{X_2}$). Third, the sum
(difference) modulus 2$\pi$ of the two angular probability integral transforms was calculated. Finally, the Rayleigh or Pycke test was applied to the vector of the
observed values of the sum (difference) of the APITs. In the case of a positive association between the random variables, the proposed independence test must be applied to
the difference of the APITs and in the case of a negative association, it must be applied to the sum of the APITs. In the case in which there is no prior indication regarding whether the association is positive or negative, one can calculate the two test statistics, one for the sum and the other for the difference of the APITs, and one can modify the minimum of the two p values in accordance with some multiple testing correction, such as the Bonferroni procedure (see Cinar and Viechtbauer, 2022 and Goeman and Solari, 2014). In practice, the user can plot the histogram of the sum (difference) of APITs and decide on the use of the Rayleigh or Pycke test in terms of the observed number of modes. In case of doubt, the Pycke test is preferred. Other omnibus tests for circular uniformity can be considered as those of Ajne, Watson and Rao (see Mardia and Jupp, 2000).

After fitting an NNTS model with $M=1$ to the sum (difference) of
APITs, the following measure of dependence can be defined:
\begin{equation}
\hat{\lambda}_{c_0}=\frac{M+1}{M}\left(1-\hat{c}_0^2 \right)=2\left(1-\hat{c}_0^2 \right)
\end{equation}
where the correction term $\frac{M+1}{M}=2$ comes from the fact that the NNTS density, in the case $M=1$, with the highest concentration around zero has a parameter vector $\underline{c}$ in which the squared norm of each of its components is equal to $\frac{1}{M+1}=\frac{1}{2}$. This implies that $\frac{1}{2} \le c_0^2 \le 1$ and $\hat{\lambda}_{c_0}$ as $M=1$ takes values in the interval $[0,1]$ with values close to zero, implying low dependence (independence), and values closer to one, further implying high dependence between the considered random variables. For independent random variables, the theoretical value of $\hat{\lambda}_{c_0}$ is equal to zero. The measure of dependence $\hat{\lambda}_{c_0}$ is particularly useful in the circular-circular and circular-linear cases.

\section{Simulation Study}

In this section, we present a simulation study to compare the power of the proposed test with that of the Wilks test and that of a test of independence based on the empirical copula.
We simulated the data from different multivariate distributions by using known parameters that define the dependence structure and known marginal densities.
We considered sample sizes of 20, 50, 100, and 200. For a given significance level $\alpha$ (10\%, 5\%, and 1\%), the powers of the tests were obtained by generating
100 samples of the specified sample size from the bivariate density by calculating the values of the test statistics for each of the 100 samples and by determining
the number of times that the test statistics considered a value that rejected the null hypothesis of independence at the given value of $\alpha$. Thus, the reported powers
of the tests considered values in the range of 0-100 and can be interpreted in terms of percentages. The Rayleigh test for circular uniformity was performed by using the
$circular$ $R$ package (Agostinelli and Lund, 2017). The Pycke test of circular uniformity was performed by using the $CircMLE$ $R$ package (Fitak and
Johnsen, 2017 and Landler et al., 2019). The $R$ package $copula$ was used to calculate the empirical copula test, and the measure of dependence $\hat{\lambda}_{c_0}$ was
obtained by fitting an NNTS model with $M=1$ by using the $R$ package $CircNNTSR$ (Fern\'andez-Dur\'an and Gregorio-Dom\'inguez, 2016).

\subsection{Circular-Linear Models}

Table \ref{Circularlinearpower} includes the powers of the proposed independence test when using a Rayleigh (ART) and Pycke (APT) circular uniformity tests with respect to those of
the Wilks (WT) and empirical copula (ECT) independence tests when simulating samples from the circular-linear model of
Johnson and Wehrly with the circular marginal density being an NNTS density with $M=3$, which is plotted in the first plot of Figure \ref{graphcircularcircularcopula}; there are three different linear models (exponential, Gaussian, and Cauchy) and a joining circular density $g$ that corresponds to an NNTS density with $M=3$ with five different values
of the parameter $c_0$ (0.7, 0.8, 0.9, 0.99, and 0.9999) to account for different degrees of association between the circular and linear random variables, as depicted in the last plot of Figure \ref{graphcircularcircularcopula}, which includes the plots of the angular joining functions for the five different values of parameter $c_0$.
The case $c_0=0.9999$ corresponds to an almost circular uniform density and to the null hypothesis of independence (refer to the last plot of Figure
\ref{graphcircularcircularcopula}). In general, the proposed ART and APT tests demonstrated significantly larger powers when compared to that of the ECT test, which was
only similar for large sample sizes and
low values of $c_0$ (0.7, 0.8, and 0.9), further representing highly dependent circular and linear random variables. The power of the WT was considerably lower when compared to that of the ART, APT and ECT tests.

\subsection{Circular-Circular Models}

For Johnson and Wehrly's circular-circular model, we used the same angular joining density and one of the marginal circular densities as that used in the
circular-linear model. Figure \ref{graphcircularcircularcopula} depicts the plots of the marginal circular densities that correspond to NNTS densities with $M=3$ and $M=2$ and angular joining densities $g$ that correspond to NNTS densities with $M=3$ for five different values of the parameter $c_0$ (0.7, 0.8, 0.9, 0.99, and 0.9999)
for the circular-circular model of Johnson and Wehrly. Similar to the results in the circular-linear model, the ART and APT demonstrated larger power values when compared to those of the WT and ECT tests. Given the multimodality of all the circular densities involved, of the two proposed independence tests, APT exhibited a larger power when compared to that of ART, which was similar only for the largest sample size of 200. Moreover, for highly dependent circular random variables ($c_0$ = 0.7, 0.8, or 0.9) and large sample sizes of 100
or 200, the power of the ECT test was similar to those of the ART and APT tests. The average values of the dependence measure $\hat{\lambda}_{c_0}$ listed in the fourth column of
Tables \ref{Circularlinearpower} and \ref{Circularcircularpower}
assumed similar values when $c_0=$0.7, 0.8, and 0.9 but these values were smaller when $c_0$=0.99 and 0.9999, further reflecting the fact that values of $c_0$ near one are associated with random variables with a very weak association.

\subsection{Linear-Linear Models}

Tables \ref{Gaussiancopulapower} and \ref{Frankcopulapower} list the powers of the different tests while simulating samples from a bivariate distribution
in which both variables are linear. In the first case, a Gaussian copula was used, and in the second case a Frank copula was used.

\subsubsection{Bivariate Gaussian Copula}

The bivariate Gaussian (normal) copula corresponds to a multivariate distribution,
which is defined as follows:
\begin{equation}
C(u_1, u_2)=\Phi_{\Gamma}(\Phi^{-1}(u_1),\Phi^{-1}(u_2))
\end{equation}
where $\Phi_{\Gamma}$ is the multivariate normal distribution with a zero mean vector and correlation matrix $\Gamma$ and $\Phi()$ is the univariate standard normal distribution function. By using an identity matrix as a correlation matrix, $\Gamma=I$, the independence copula, $C(u_1,u_2)=u_1u_2$ is obtained.

Table \ref{Gaussiancopulapower} compares the powers of the proposed independence test when using Rayleigh (ART) and Pycke (APT) circular uniformity tests with respect to those of
the Wilks (WT) and empirical copula (ECT) independence tests when using simulated samples from a bivariate linear-linear distribution with a Gaussian copula and three different
cases of marginal distributions following Herwatz and Maxand (2020): exponential, Gaussian, and Cauchy. For the Gaussian copula, we considered five different values of the correlation coefficient $\rho$ (0, 0.25, 0.5, 0.75, and 0.99). When the correlation coefficient is equal to zero, it corresponds to the null
independence hypothesis, and the reported powers correspond to the sizes of the tests expected to be similar to the significance levels $\alpha$ (10\%, 5\%,
and 1\%). In general terms, both WT and ECT have higher powers than those of the proposed ART and APT tests. However, for large sample sizes and high values of
the correlation coefficient (0.75 and 0.99), the ART and APT tests have powers similar to those of the WT and ECT tests. As expected, for the exponential and Cauchy marginals, the power of the WT is reduced when compared to that of the Gaussian marginal case for which it was designed; that is, the power of the WT deteriorates for
marginals that are not Gaussian. In the case of Gaussian marginals, the WT for large sample sizes has high power, and in some cases, its power is larger than the
ECT power. In terms of the sizes of all the tests, it appears that all tests have approximately the correct sizes, given that a total of 100 samples were used. The fourth column of Table \ref{Gaussiancopulapower} includes the averages of the values of the dependence measure $\hat{\lambda}_{c_0}$ which, as expected, increase as the value of the correlation coefficient $\rho$ increases.

\subsubsection{Bivariate Frank Copula}

The Frank bivariate copula is defined as follows:
\begin{equation}
C(u,v)=-\frac{1}{\varphi}\ln\left(1 + \frac{(e^{-\varphi u}-1)(e^{-\varphi v}-1)}{e^{-\varphi}-1} \right)
\end{equation}
where $u, v \in (0,1]$. Parameter $\varphi$ assumes values in the interval $(0,\infty)$ and in the limit $\varphi=0$, the Frank copula corresponds to the independence
copula. In Table \ref{Frankcopulapower}, the Gaussian copula of the previous study is replaced by the Frank copula by using five different values of the dependence parameter
$\varphi$ (0, 5, 10, 15, and 50). For the case of independence obtained in the limit when $\varphi=0$, the powers are also obtained when the limit is considered as $\varphi=0$. In general, in Table \ref{Frankcopulapower}, we obtained the same conclusions as those obtained in Table \ref{Gaussiancopulapower}, further indicating that the
characteristics of the proposed tests, ART and APT, are more suitable for the circular-circular and circular-linear cases than for the linear-linear case.

\section{Application to Real Circular-Circular and Circular-Linear Data}

\subsection{Test of Bivariate Independence}

\subsubsection{Circular-Linear Real Examples}

Figure \ref{graphcircularcircularexamples} depicts the scatterplots of the considered real examples.
We applied the proposed independence test to the circular-linear data on wind direction (circular variable) and ozone concentration (linear variable) originally analyzed
by Johnson and Wehrly (1977) and later included them as dataset B.18 in Fisher (1993). A total of 19 measurements were taken at a weather station
in Milwaukee at 6 o'clock in the morning every fourth day starting on April 18 and ending on June 29, 1975.
The scatterplot of these data is included in the top left plot of Figure \ref{graphcircularcircularexamples}, which presents the values
of the wind direction and ozone concentration, further indicating a possible positive association between the circular and linear variables while considering the periodicity of the
circular random variable. By applying the Pycke and Rayleigh circular uniformity tests to the difference in the angular probability transforms of the circular and linear
variables, we obtained p values of 0.0133 and 0.0077, respectively, thus rejecting the null hypothesis of independence (uniformity) at a 5\% significance level for the wind direction and ozone concentration. The value of the dependence measure $\hat{\lambda}_{c_0}$ was calculated to be 0.4383. Fisher (1993) reached the same conclusion by considering an expected sine-wave functional form for the conditional expected
value of ozone concentration given the wind direction.

A second example analyzed by Fisher (1993) is a dataset on the directions and distances traveled by 31 small blue periwinkles after undergoing transplantation from their normal living place. The top-right plot depicted in Figure \ref{graphcircularcircularexamples} includes the scatterplot for this dataset, further indicating a possible negative association between
the direction and travel distance. When applying the uniformity test to the sum of the angular probability transforms of the direction and distance, a p value of 0.0087
for the Pycke test and a p value of 0.0096 for the Rayleigh test were obtained, which means we rejected independence in accordance with the results obtained by Fisher (1993) when fitting a
circular-linear regression model with von Mises errors with non constant dispersion. The dependence measure $\hat{\lambda}_{c_0}$ was calculated to be 0.2668.

\subsubsection{Circular-Circular Real Examples}

The first example in the circular-circular test of independence corresponds to pairs of wind directions measured at a weather monitoring station at Milwaukee. The measurements
were taken at 6:00 and 12:00 o'clock for 21 consecutive days and were originally included in Johnson and Wehrly (1977). The bottom-left plot depicted in Figure
\ref{graphcircularcircularexamples} includes a scatterplot of the pairs of wind directions, which indicates a possible positive association between the two angles. Fisher (1993)
listed this dataset as the B.21 dataset, and the main conclusion of Fisher (1993) was that there exists a strong positive association between the wind directions when
applying a hypothesis test based on a circular-circular correlation coefficient. When applying the proposed methodology to the difference of the angular
probability transforms, a p value of 0.0148 for the Pycke test and a p value of 0.0075 for the Rayleigh test were obtained, which means we rejected the null hypothesis of independence
between the two angles at a 5\% significance level. The dependence measure $\hat{\lambda}_{c_0}$ was calculated to be one.

The second example corresponds to 233 pairs ($\phi$,$\psi$) of dihedral angles in alanine-alanine-alanine segments of proteins that were originally analyzed by
Fern\'{a}ndez-Dur\'{a}n (2007) by using bivariate
NNTS models demonstrating the dependence between the two angles. The scatterplot of these two angles is included in the bottom-right plot of Figure
\ref{graphcircularcircularexamples}, and the type of association between the two angles, whether it is positive or negative is not evident. When applying the independence test to the
difference of the angular probability transforms of the two angles, we obtained p values of 0.0014 and 0.0064 for the Rayleigh and Pycke circular uniformity tests,
respectively, thus clearly rejecting the null hypothesis of independence in favor of a positive association.

\section{Conclusions}

By using the result that the sum of independent circular uniform random variables is circular and uniformly distributed,
a general test for independence based on the angular integral probability transform was developed. We demonstrated its use, particularly when at least one of the variables
is an angle, that is, a circular random variable, further implying that testing for independence in the circular case is equivalent to testing for circular uniformity. From the
simulation study presented in this paper, it is clear that the proposed
test is particularly useful for bivariate cases of circular-circular and
circular-linear pairs of random variables with more power than the Wilks and empirical copula independence tests. We reached this conclusion by simulating samples from NNTS densities, in which the degree of closeness to the circular uniform distribution was under control. Although the proposed independence test can be applied to the linear-linear case, its power is smaller than that of the commonly used independence tests converging to the power of the common independence tests when the sample size increases. The proposed independence
test can be put in practice, and it demonstrated superior performance when at least one circular random variable is among the random variables used to test for independence.
Additionally, a new measure of dependence was introduced based on the fitting of an NNTS density by considering $M=1$ to the sum (difference) of the angular probability integral transforms.

\section*{Acknowledgments}
We express our sincere gratitude to the Asociaci\'on Mexicana de Cultura, A.C. for their support.

\newpage

\renewcommand{\baselinestretch}{1.00}
\begin{table}[t]
\begin{center}
\scalebox{0.6}{
\begin{tabular}{|c|c|c|c|cccc|cccc|cccc|}
\hline
          &    &       & $\hat{\lambda}_{c0}=$           & \multicolumn{4}{|c|}{$\alpha=10\%$} & \multicolumn{4}{|c|}{$\alpha=5\%$} & \multicolumn{4}{|c|}{$\alpha=1\%$} \\
Marginals & SS & $c_0$ & $2(1-\hat{c}_0^2)$ & RT & PT & WT & ECT                  & RT & PT & WT & ECT                 & RT & PT & WT & ECT \\
\hline
$\Theta \sim NNTS(M=3) $    & 20  & 0.7    & 0.61 & 91  & 89  & 29 & 43  & 89  & 88  & 22 & 32  & 68  & 64  & 8  & 3 \\
$X \sim Exp(1)$             & 20  & 0.8    & 0.66 & 88  & 85  & 26 & 48  & 83  & 79  & 22 & 39  & 67  & 64  & 6  & 5 \\
                            & 20  & 0.9    & 0.62 & 86  & 85  & 39 & 47  & 83  & 80  & 21 & 35  & 58  & 61  & 8  & 11 \\
                            & 20  & 0.99   & 0.22 & 34  & 35  & 18 & 11  & 28  & 26  & 10 & 9   & 10  & 7   & 3  & 2 \\
                            & 20  & 0.9999 & 0.15 & 13  & 10  & 12 & 10  & 3   & 5   & 7  & 7   & 1   & 1   & 1  & 2 \\
                            & 50  & 0.7    & 0.46 & 100 & 100 & 33 & 79  & 100 & 99  & 20 & 57  & 97  & 97  & 11 & 28 \\
                            & 50  & 0.8    & 0.51 & 100 & 100 & 33 & 86  & 100 & 100 & 18 & 61  & 99  & 98  & 6  & 29 \\
                            & 50  & 0.9    & 0.5  & 100 & 100 & 43 & 89  & 100 & 100 & 34 & 72  & 99  & 98  & 9  & 48 \\
                            & 50  & 0.99   & 0.15 & 71  & 81  & 21 & 40  & 64  & 70  & 13 & 20  & 40  & 41  & 4  & 3 \\
                            & 50  & 0.9999 & 0.04 & 12  & 10  & 7  & 10  & 6   & 6   & 3  & 4   & 1   & 2   & 0  & 2 \\
                            & 100 & 0.7    & 0.45 & 100 & 100 & 51 & 100 & 100 & 100 & 39 & 100 & 100 & 100 & 19 & 89 \\
                            & 100 & 0.8    & 0.49 & 100 & 100 & 47 & 100 & 100 & 100 & 30 & 99  & 100 & 100 & 16 & 94 \\
                            & 100 & 0.9    & 0.47 & 100 & 100 & 72 & 100 & 100 & 100 & 56 & 100 & 100 & 100 & 26 & 90 \\
                            & 100 & 0.99   & 0.12 & 90  & 95  & 31 & 55  & 82  & 92  & 22 & 41  & 66  & 80  & 7  & 18 \\
                            & 100 & 0.9999 & 0.02 & 10  & 7   & 11 & 9   & 5   & 4   & 4  & 4   & 1   & 0   & 1  & 2 \\
                            & 200 & 0.7    & 0.44 & 100 & 100 & 71 & 100 & 100 & 100 & 64 & 100 & 100 & 100 & 42 & 100 \\
                            & 200 & 0.8    & 0.49 & 100 & 100 & 65 & 100 & 100 & 100 & 58 & 100 & 100 & 100 & 31 & 100 \\
                            & 200 & 0.9    & 0.45 & 100 & 100 & 86 & 100 & 100 & 100 & 76 & 100 & 100 & 100 & 56 & 100 \\
                            & 200 & 0.99   & 0.11 & 100 & 100 & 39 & 97  & 100 & 100 & 31 & 79  & 99  & 100 & 13 & 55 \\
                            & 200 & 0.9999 & 0.01 & 11  & 14  & 10 & 12  & 9   & 7   & 3  & 8   & 3   & 1   & 0  & 0 \\
\hline
$\Theta \sim NNTS(M=3) $    & 20  & 0.7    & 0.62 & 89  & 86  & 13 & 36  & 79  & 80  & 7  & 25  & 55  & 52  & 2  & 4 \\
$X \sim N(0,1)$             & 20  & 0.8    & 0.63 & 89  & 85  & 17 & 39  & 78  & 76  & 8  & 23  & 53  & 52  & 6  & 6 \\
                            & 20  & 0.9    & 0.64 & 94  & 88  & 20 & 48  & 86  & 78  & 8  & 37  & 62  & 57  & 1  & 9 \\
                            & 20  & 0.99   & 0.32 & 38  & 30  & 17 & 23  & 28  & 19  & 11 & 19  & 11  & 10  & 1  & 8 \\
                            & 20  & 0.9999 & 0.16 & 13  & 18  & 18 & 14  & 7   & 8   & 10 & 12  & 1   & 2   & 3  & 3 \\
                            & 50  & 0.7    & 0.56 & 100 & 100 & 19 & 92  & 100 & 100 & 8  & 77  & 99  & 99  & 1  & 34 \\
                            & 50  & 0.8    & 0.59 & 100 & 100 & 12 & 89  & 100 & 100 & 7  & 78  & 99  & 99  & 2  & 33 \\
                            & 50  & 0.9    & 0.49 & 100 & 100 & 29 & 85  & 100 & 100 & 18 & 64  & 98  & 99  & 7  & 30 \\
                            & 50  & 0.99   & 0.13 & 59  & 63  & 21 & 27  & 52  & 55  & 13 & 11  & 30  & 33  & 5  & 5 \\
                            & 50  & 0.9999 & 0.04 & 14  & 12  & 8  & 10  & 5   & 6   & 4  & 5   & 2   & 2   & 1  & 1 \\
                            & 100 & 0.7    & 0.43 & 100 & 100 & 12 & 100 & 100 & 100 & 7  & 100 & 100 & 100 & 2  & 95 \\
                            & 100 & 0.8    & 0.5  & 100 & 100 & 14 & 100 & 100 & 100 & 10 & 100 & 100 & 100 & 0  & 97 \\
                            & 100 & 0.9    & 0.48 & 100 & 100 & 44 & 100 & 100 & 100 & 29 & 98  & 100 & 100 & 13 & 93 \\
                            & 100 & 0.99   & 0.13 & 91  & 97  & 21 & 64  & 86  & 91  & 14 & 36  & 66  & 76  & 4  & 23 \\
                            & 100 & 0.9999 & 0.02 & 10  & 14  & 7  & 13  & 5   & 7   & 4  & 3   & 1   & 1   & 1  & 2 \\
                            & 200 & 0.7    & 0.44 & 100 & 100 & 18 & 100 & 100 & 100 & 9  & 100 & 100 & 100 & 4  & 100 \\
                            & 200 & 0.8    & 0.48 & 100 & 100 & 16 & 100 & 100 & 100 & 10 & 100 & 100 & 100 & 3  & 100 \\
                            & 200 & 0.9    & 0.46 & 100 & 100 & 64 & 100 & 100 & 100 & 56 & 100 & 100 & 100 & 31 & 100 \\
                            & 200 & 0.99   & 0.11 & 100 & 100 & 30 & 92  & 100 & 100 & 15 & 81  & 97  & 100 & 8  & 50 \\
                            & 200 & 0.9999 & 0.01 & 11  & 7   & 9  & 4   & 6   & 3   & 8  & 1   & 0   & 0   & 3  & 0 \\
\hline
$\Theta \sim NNTS(M=3) $    & 20  & 0.7    & 0.52 & 81  & 75  & 17 & 28  & 73  & 66  & 13 & 14  & 46  & 46  & 6  & 6 \\
$X \sim Cauchy(0,1)$        & 20  & 0.8    & 0.58 & 82  & 84  & 15 & 40  & 78  & 72  & 9  & 22  & 53  & 50  & 1  & 5 \\
                            & 20  & 0.9    & 0.6  & 81  & 79  & 22 & 34  & 70  & 71  & 14 & 22  & 56  & 52  & 3  & 9 \\
                            & 20  & 0.99   & 0.26 & 34  & 37  & 12 & 21  & 22  & 25  & 8  & 9   & 7   & 10  & 1  & 4 \\
                            & 20  & 0.9999 & 0.15 & 12  & 16  & 12 & 15  & 6   & 10  & 10 & 6   & 2   & 6   & 3  & 0 \\
                            & 50  & 0.7    & 0.49 & 100 & 100 & 26 & 84  & 100 & 100 & 17 & 68  & 99  & 99  & 1  & 35 \\
                            & 50  & 0.8    & 0.5  & 100 & 100 & 15 & 89  & 100 & 100 & 8  & 80  & 99  & 98  & 0  & 35 \\
                            & 50  & 0.9    & 0.48 & 100 & 100 & 13 & 88  & 100 & 100 & 6  & 67  & 99  & 99  & 2  & 32 \\
                            & 50  & 0.99   & 0.13 & 62  & 69  & 13 & 28  & 52  & 56  & 8  & 19  & 26  & 26  & 2  & 5 \\
                            & 50  & 0.9999 & 0.05 & 12  & 12  & 15 & 9   & 4   & 5   & 9  & 3   & 1   & 2   & 3  & 1 \\
                            & 100 & 0.7    & 0.45 & 100 & 100 & 13 & 100 & 100 & 100 & 10 & 100 & 100 & 100 & 4  & 89 \\
                            & 100 & 0.8    & 0.49 & 100 & 100 & 14 & 100 & 100 & 100 & 10 & 99  & 100 & 100 & 3  & 89 \\
                            & 100 & 0.9    & 0.47 & 100 & 100 & 12 & 100 & 100 & 100 & 7  & 97  & 100 & 100 & 2  & 83 \\
                            & 100 & 0.99   & 0.12 & 97  & 94  & 12 & 52  & 88  & 94  & 10 & 34  & 74  & 84  & 3  & 15 \\
                            & 100 & 0.9999 & 0.02 & 9   & 14  & 7  & 7   & 4   & 5   & 6  & 5   & 0   & 1   & 2  & 0 \\
                            & 200 & 0.7    & 0.43 & 100 & 100 & 17 & 100 & 100 & 100 & 11 & 100 & 100 & 100 & 0  & 100 \\
                            & 200 & 0.8    & 0.48 & 100 & 100 & 17 & 100 & 100 & 100 & 13 & 100 & 100 & 100 & 1  & 100 \\
                            & 200 & 0.9    & 0.46 & 100 & 100 & 15 & 100 & 100 & 100 & 7  & 100 & 100 & 100 & 1  & 100 \\
                            & 200 & 0.99   & 0.1  & 100 & 100 & 13 & 92  & 99  & 100 & 5  & 84  & 97  & 100 & 3  & 56 \\
                            & 200 & 0.9999 & 0.01 & 14  & 13  & 10 & 7   & 7   & 5   & 8  & 2   & 2   & 1   & 1  & 1 \\
\hline
\end{tabular}}
\renewcommand{\baselinestretch}{1}
\caption{NNTS angular joining density $g$ circular-linear power study: The powers of the proposed test implemented by using the Rayleigh (ART) and Pycke (APT) circular uniformity
tests, the Wilks (WT) and the empirical copula (ECT) tests are compared when simulating 100 times samples of sizes 20, 50, 100 and 200 from a Johnson and Wehrly
circular-linear density function constructed from an NNTS angular joining density with $M=3$, an NNTS marginal density function with $M=3$ (see Figure \ref{graphcircularcircularcopula}) and three different linear
marginals (exponential, Gaussian and Cauchy). The NNTS angular joining density is defined with five different values of the parameter $c_0$ (0.7, 0.8, 0.9, 0.99
and 0.9999). The case with $c_0=1$ corresponds to the null independence model.
\label{Circularlinearpower} }
\end{center}
\end{table}

\newpage

\renewcommand{\baselinestretch}{1.00}
\begin{table}[t]
\begin{center}
\scalebox{0.6}{
\begin{tabular}{|c|c|c|c|cccc|cccc|cccc|}
\hline
          &    &       & $\hat{\lambda}_{c0}=$           & \multicolumn{4}{|c|}{$\alpha=10\%$} & \multicolumn{4}{|c|}{$\alpha=5\%$} & \multicolumn{4}{|c|}{$\alpha=1\%$} \\
Marginals & SS & $c_0$ & $2(1-\hat{c}_0^2)$ & RT & PT & WT & ECT                  & RT & PT & WT & ECT                 & RT & PT & WT & ECT \\
\hline
$\Theta_1 \sim NNTS(M_1=3) $    & 20  & 0.7    & 0.63 & 87  & 80  & 16 & 38  & 78  & 73  & 11 & 29  & 59  & 61  & 1  & 11 \\
$\Theta_2 \sim NNTS(M_2=2) $    & 20  & 0.8    & 0.6  & 84  & 81  & 13 & 41  & 72  & 74  & 9  & 28  & 57  & 57  & 3  & 10 \\
                                & 20  & 0.9    & 0.59 & 81  & 79  & 20 & 36  & 76  & 67  & 15 & 28  & 46  & 49  & 5  & 7 \\
                                & 20  & 0.99   & 0.25 & 34  & 33  & 19 & 15  & 26  & 23  & 13 & 12  & 10  & 8   & 7  & 4 \\
                                & 20  & 0.9999 & 0.13 & 11  & 13  & 12 & 10  & 7   & 7   & 7  & 10  & 1   & 0   & 1  & 1 \\
                                & 50  & 0.7    & 0.45 & 100 & 100 & 16 & 88  & 99  & 99  & 7  & 63  & 99  & 98  & 1  & 41 \\
                                & 50  & 0.8    & 0.5  & 100 & 100 & 15 & 91  & 100 & 100 & 5  & 70  & 98  & 98  & 0  & 38 \\
                                & 50  & 0.9    & 0.49 & 100 & 100 & 28 & 92  & 100 & 100 & 14 & 70  & 100 & 100 & 6  & 34 \\
                                & 50  & 0.99   & 0.14 & 70  & 73  & 25 & 36  & 55  & 63  & 16 & 17  & 36  & 36  & 7  & 6 \\
                                & 50  & 0.9999 & 0.05 & 17  & 14  & 9  & 12  & 6   & 6   & 4  & 3   & 0   & 1   & 3  & 1 \\
                                & 100 & 0.7    & 0.42 & 100 & 100 & 13 & 100 & 100 & 100 & 9  & 100 & 100 & 100 & 3  & 92 \\
                                & 100 & 0.8    & 0.47 & 100 & 100 & 17 & 100 & 100 & 100 & 9  & 100 & 100 & 100 & 2  & 93 \\
                                & 100 & 0.9    & 0.46 & 100 & 100 & 38 & 100 & 100 & 100 & 28 & 100 & 100 & 100 & 10 & 96 \\
                                & 100 & 0.99   & 0.12 & 90  & 95  & 32 & 60  & 83  & 92  & 23 & 41  & 64  & 85  & 10 & 15 \\
                                & 100 & 0.9999 & 0.02 & 9   & 14  & 17 & 8   & 4   & 7   & 7  & 1   & 3   & 1   & 1  & 0 \\
                                & 200 & 0.7    & 0.43 & 100 & 100 & 25 & 100 & 100 & 100 & 19 & 100 & 100 & 100 & 6  & 100 \\
                                & 200 & 0.8    & 0.48 & 100 & 100 & 18 & 100 & 100 & 100 & 12 & 100 & 100 & 100 & 6  & 100 \\
                                & 200 & 0.9    & 0.45 & 100 & 100 & 51 & 100 & 100 & 100 & 45 & 100 & 100 & 100 & 21 & 100 \\
                                & 200 & 0.99   & 0.1  & 100 & 100 & 42 & 86  & 100 & 100 & 31 & 80  & 99  & 99  & 15 & 40 \\
                                & 200 & 0.9999 & 0.01 & 8   & 10  & 12  & 9   & 4   & 6   & 6 & 4   & 0   & 0   & 2  & 1 \\
\hline
\end{tabular}}
\renewcommand{\baselinestretch}{1}
\caption{NNTS angular joining density circular-circular power study: The powers of the proposed test implemented by using the Rayleigh (ART) and Pycke (APT) circular
uniformity tests, the Wilks (WT) and the empirical copula (ECT) tests are compared when simulating 100 times samples of sizes 20, 50, 100 and 200 from a Johnson and Wehrly
circular-circular density function constructed from an NNTS angular joining density with $M=3$ and NNTS marginal density functions with $M_1=3$ and $M_2=2$. The NNTS
angular joining density is defined with five different values of the dependence parameter $c_0$ (0.7, 0.8, 0.9, 0.99 and 0.9999). The case with $c_0=1$ corresponds to the
null independence model. The plots of the NNTS angular joining density and NNTS circular marginal densities are shown in Figure \ref{graphcircularcircularcopula}.
\label{Circularcircularpower} }
\end{center}
\end{table}

\newpage

\renewcommand{\baselinestretch}{1.00}
\begin{table}[t]
\begin{center}
\scalebox{0.6}{
\begin{tabular}{|c|c|c|c|cccc|cccc|cccc|}
\hline
          &    &       & $\hat{\lambda}_{c0}=$           & \multicolumn{4}{|c|}{$\alpha=10\%$} & \multicolumn{4}{|c|}{$\alpha=5\%$} & \multicolumn{4}{|c|}{$\alpha=1\%$} \\
Marginals & SS & $\rho$ & $2(1-\hat{c}_0^2)$ & ART & APT & WT & ECT                  & ART & APT & WT & ECT                 & ART & APT & WT & ECT \\
\hline
$X_1 \sim Exp(1)$       & 20  & 0.99 & 1    & 100 & 100 & 100 & 100 & 100 & 100 & 100 & 100 & 100 & 100 & 100 & 100 \\
$X_2 \sim Exp(2)$       & 20  & 0.75 & 0.57 & 75  & 69  & 98  & 97  & 68  & 60  & 92  & 92  & 50  & 45  & 83  & 75 \\
                        & 20  & 0.5  & 0.25 & 38  & 36  & 64  & 60  & 24  & 25  & 51  & 47  & 9   & 7   & 40  & 24 \\
                        & 20  & 0.25 & 0.15 & 17  & 20  & 33  & 29  & 10  & 8   & 24  & 15  & 3   & 4   & 9   & 3 \\
                        & 20  & 0    & 0.14 & 10  & 14  & 12  & 10  & 7   & 9   & 8   & 5   & 0   & 0   & 2   & 3 \\
                        & 50  & 0.99 & 1    & 100 & 100 & 100 & 100 & 100 & 100 & 100 & 100 & 100 & 100 & 100 & 100 \\
                        & 50  & 0.75 & 0.48 & 99  & 98  & 100 & 100 & 99  & 96  & 100 & 100 & 91  & 89  & 100 & 100 \\
                        & 50  & 0.5  & 0.14 & 57  & 51  & 93  & 94  & 46  & 38  & 87  & 90  & 23  & 21  & 71  & 85 \\
                        & 50  & 0.25 & 0.06 & 18  & 17  & 52  & 42  & 10  & 7   & 42  & 28  & 2   & 1   & 14  & 15 \\
                        & 50  & 0    & 0.05 & 7   & 6   & 11  & 8   & 1   & 3   & 6   & 4   & 1   & 0   & 4   & 0 \\
                        & 100 & 0.99 & 1    & 100 & 100 & 100 & 100 & 100 & 100 & 100 & 100 & 100 & 100 & 100 & 100 \\
                        & 100 & 0.75 & 0.43 & 100 & 100 & 100 & 100 & 100 & 100 & 100 & 100 & 99  & 99  & 100 & 100 \\
                        & 100 & 0.5  & 0.11 & 79  & 76  & 97  & 99  & 72  & 61  & 96  & 99  & 50  & 43  & 94  & 98 \\
                        & 100 & 0.25 & 0.03 & 32  & 30  & 71  & 69  & 20  & 15  & 58  & 55  & 4   & 5   & 37  & 36 \\
                        & 100 & 0    & 0.02 & 8   & 12  & 13  & 10  & 4   & 3   & 10  & 5   & 1   & 1   & 4   & 1 \\
                        & 200 & 0.99 & 1    & 100 & 100 & 100 & 100 & 100 & 100 & 100 & 100 & 100 & 100 & 100 & 100 \\
                        & 200 & 0.75 & 0.4  & 100 & 100 & 100 & 100 & 100 & 100 & 100 & 100 & 100 & 100 & 100 & 100 \\
                        & 200 & 0.5  & 0.08 & 99  & 98  & 100 & 100 & 97  & 97  & 100 & 100 & 88  & 82  & 100 & 100 \\
                        & 200 & 0.25 & 0.02 & 33  & 27  & 90  & 94  & 21  & 24  & 86  & 90  & 10  & 7   & 63  & 74 \\
                        & 200 & 0    & 0.01 & 10  & 13  & 13  & 6   & 6   & 7   & 7   & 1   & 1   & 1   & 1   & 0 \\
\hline
$X_1 \sim N(0,1)$       & 20  & 0.99 & 1    & 100 & 100 & 100 & 100 & 100 & 100 & 100 & 100 & 100 & 100 & 100 & 100 \\
$X_2 \sim N(0,1)$       & 20  & 0.75 & 0.6  & 81  & 75  & 99  & 97  & 70  & 64  & 99  & 93  & 50  & 37  & 94  & 72 \\
                        & 20  & 0.5  & 0.26 & 27  & 21  & 73  & 66  & 19  & 13  & 65  & 54  & 7   & 4   & 42  & 25 \\
                        & 20  & 0.25 & 0.14 & 13  & 10  & 38  & 21  & 4   & 7   & 25  & 12  & 1   & 2   & 11  & 2 \\
                        & 20  & 0    & 0.15 & 13  & 8   & 15  & 7   & 4   & 6   & 8   & 5   & 1   & 1   & 2   & 2 \\
                        & 50  & 0.99 & 1    & 100 & 100 & 100 & 100 & 100 & 100 & 100 & 100 & 100 & 100 & 100 & 100 \\
                        & 50  & 0.75 & 0.49 & 99  & 99  & 100 & 100 & 98  & 97  & 100 & 100 & 96  & 92  & 100 & 100 \\
                        & 50  & 0.5  & 0.12 & 59  & 51  & 99  & 95  & 40  & 40  & 98  & 92  & 26  & 22  & 92  & 75 \\
                        & 50  & 0.25 & 0.05 & 19  & 15  & 58  & 45  & 8   & 11  & 47  & 32  & 1   & 1   & 28  & 13 \\
                        & 50  & 0    & 0.04 & 5   & 8   & 13  & 10  & 1   & 5   & 6   & 5   & 1   & 1   & 1   & 3 \\
                        & 100 & 0.99 & 1    & 100 & 100 & 100 & 100 & 100 & 100 & 100 & 100 & 100 & 100 & 100 & 100 \\
                        & 100 & 0.75 & 0.44 & 100 & 100 & 100 & 100 & 100 & 100 & 100 & 100 & 100 & 100 & 100 & 100 \\
                        & 100 & 0.5  & 0.1  & 81  & 75  & 100 & 100 & 69  & 66  & 100 & 100 & 47  & 44  & 99  & 100 \\
                        & 100 & 0.25 & 0.03 & 23  & 18  & 80  & 71  & 12  & 11  & 75  & 56  & 3   & 4   & 57  & 33 \\
                        & 100 & 0    & 0.02 & 4   & 6   & 9   & 5   & 1   & 5   & 4   & 1   & 0   & 0   & 2   & 0 \\
                        & 200 & 0.99 & 1    & 100 & 100 & 100 & 100 & 100 & 100 & 100 & 100 & 100 & 100 & 100 & 100 \\
                        & 200 & 0.75 & 0.4  & 100 & 100 & 100 & 100 & 100 & 100 & 100 & 100 & 100 & 100 & 100 & 100 \\
                        & 200 & 0.5  & 0.09 & 98  & 97  & 100 & 100 & 96  & 92  & 100 & 100 & 88  & 83  & 100 & 100 \\
                        & 200 & 0.25 & 0.02 & 41  & 28  & 98  & 93  & 22  & 19  & 93  & 90  & 9   & 8   & 83  & 79 \\
                        & 200 & 0    & 0.01 & 11  & 11  & 10  & 11  & 8   & 5   & 5   & 5   & 0   & 0   & 0   & 2 \\
\hline
$X_1 \sim Cauchy(0,1)$  & 20  & 0.99 & 1    & 100 & 100 & 100 & 100 & 100 & 100 & 100 & 100 & 100 & 100 & 100 & 100 \\
$X_2 \sim Cauchy(0,1)$  & 20  & 0.75 & 0.57 & 75  & 69  & 85  & 97  & 68  & 60  & 77  & 92  & 50  & 45  & 60  & 80 \\
                        & 20  & 0.5  & 0.25 & 38  & 36  & 42  & 60  & 24  & 25  & 35  & 47  & 9   & 7   & 23  & 26 \\
                        & 20  & 0.25 & 0.15 & 17  & 20  & 18  & 30  & 10  & 8   & 15  & 14  & 3   & 4   & 12  & 6 \\
                        & 20  & 0    & 0.14 & 10  & 14  & 16  & 10  & 7   & 9   & 13  & 5   & 0   & 0   & 7   & 4 \\
                        & 50  & 0.99 & 1    & 100 & 100 & 100 & 100 & 100 & 100 & 100 & 100 & 100 & 100 & 100 & 100 \\
                        & 50  & 0.75 & 0.48 & 99  & 98  & 93  & 100 & 99  & 96  & 90  & 100 & 91  & 89  & 81  & 100 \\
                        & 50  & 0.5  & 0.14 & 57  & 51  & 55  & 94  & 46  & 38  & 43  & 90  & 23  & 21  & 31  & 85 \\
                        & 50  & 0.25 & 0.06 & 18  & 17  & 18  & 42  & 10  & 7   & 15  & 28  & 2   & 1   & 9   & 15 \\
                        & 50  & 0    & 0.05 & 7   & 6   & 8   & 8   & 1   & 3   & 7   & 4   & 1   & 0   & 5   & 0 \\
                        & 100 & 0.99 & 1    & 100 & 100 & 100 & 100 & 100 & 100 & 100 & 100 & 100 & 100 & 100 & 100 \\
                        & 100 & 0.75 & 0.43 & 100 & 100 & 96  & 100 & 100 & 100 & 95  & 100 & 99  & 99  & 88  & 100 \\
                        & 100 & 0.5  & 0.11 & 79  & 76  & 63  & 99  & 72  & 61  & 53  & 99  & 50  & 43  & 40  & 98 \\
                        & 100 & 0.25 & 0.03 & 32  & 30  & 21  & 69  & 20  & 15  & 16  & 55  & 4   & 5   & 11  & 36 \\
                        & 100 & 0    & 0.02 & 8   & 12  & 10  & 10  & 4   & 3   & 6   & 5   & 1   & 1   & 2   & 1 \\
                        & 200 & 0.99 & 1    & 100 & 100 & 100 & 100 & 100 & 100 & 100 & 100 & 100 & 100 & 100 & 100 \\
                        & 200 & 0.75 & 0.4  & 100 & 100 & 96  & 100 & 100 & 100 & 95  & 100 & 100 & 100 & 89  & 100 \\
                        & 200 & 0.5  & 0.08 & 99  & 98  & 67  & 100 & 97  & 97  & 62  & 100 & 88  & 82  & 44  & 100 \\
                        & 200 & 0.25 & 0.02 & 33  & 27  & 19  & 94  & 21  & 24  & 13  & 90  & 10  & 7   & 10  & 74 \\
                        & 200 & 0    & 0.01 & 10  & 13  & 5   & 6   & 6   & 7   & 5   & 1   & 1   & 1   & 2   & 0 \\
\hline
\end{tabular}}
\renewcommand{\baselinestretch}{1}
\caption{Gaussian copula linear-linear power study: The powers of the proposed test implemented by using the Rayleigh (ART) and Pycke (APT) circular uniformity tests, the
Wilks (WT) and the empirical copula (ECT) tests are compared when simulating 100 times samples of sizes 20, 50, 100 and 200 from a linear-linear density function
constructed from a Gaussian copula and three different marginals (exponential, Gaussian and Cauchy). The Gaussian copula is defined with an equicorrelated correlation
matrix with five different common correlation values of 0, 0.25, 0.5, 0.75 and 0.99. The case with a common correlation equal to zero corresponds to the null independence
model.
\label{Gaussiancopulapower} } 
\end{center}
\end{table}

\renewcommand{\baselinestretch}{1.00}
\begin{table}[t]
\begin{center}
\scalebox{0.6}{
\begin{tabular}{|c|c|c|c|cccc|cccc|cccc|}
\hline
          &    &       & $\hat{\lambda}_{c0}=$           & \multicolumn{4}{|c|}{$\alpha=10\%$} & \multicolumn{4}{|c|}{$\alpha=5\%$} & \multicolumn{4}{|c|}{$\alpha=1\%$} \\
Marginals & SS & $\varphi$ & $2(1-\hat{c}_0^2)$ & ART & APT & WT & ECT                  & ART & APT & WT & ECT                 & ART & APT & WT & ECT \\
\hline
$X_1 \sim Exp(1)$       & 20  & 50 & 1    & 100 & 100 & 100 & 100 & 100 & 100 & 100 & 100 & 100 & 100 & 100 & 100 \\
$X_2 \sim Exp(2)$       & 20  & 15 & 0.95 & 100 & 100 & 100 & 100 & 100 & 99  & 100 & 100 & 99  & 97  & 98  & 100 \\
                        & 20  & 10 & 0.83 & 97  & 95  & 98  & 100 & 96  & 90  & 97  & 99  & 88  & 78  & 92  & 98 \\
                        & 20  & 5  & 0.4  & 55  & 48  & 72  & 89  & 46  & 41  & 62  & 77  & 31  & 19  & 38  & 59 \\
                        & 20  & 0  & 0.14 & 14  & 10  & 4   & 6   & 4   & 2   & 1   & 3   & 1   & 0   & 0   & 0 \\
                        & 50  & 50 & 1    & 100 & 100 & 100 & 100 & 100 & 100 & 100 & 100 & 100 & 100 & 100 & 100 \\
                        & 50  & 15 & 0.99 & 100 & 100 & 100 & 100 & 100 & 100 & 100 & 100 & 100 & 100 & 100 & 100 \\
                        & 50  & 10 & 0.9  & 100 & 100 & 100 & 100 & 100 & 100 & 100 & 100 & 100 & 100 & 100 & 100 \\
                        & 50  & 5  & 0.34 & 97  & 89  & 95  & 100 & 94  & 84  & 94  & 99  & 81  & 67  & 83  & 99 \\
                        & 50  & 0  & 0.05 & 14  & 12  & 3   & 13  & 6   & 7   & 2   & 7   & 1   & 2   & 0   & 1 \\
                        & 100 & 50 & 1    & 100 & 100 & 100 & 100 & 100 & 100 & 100 & 100 & 100 & 100 & 100 & 100 \\
                        & 100 & 15 & 1    & 100 & 100 & 100 & 100 & 100 & 100 & 100 & 100 & 100 & 100 & 100 & 100 \\
                        & 100 & 10 & 0.89 & 100 & 100 & 100 & 100 & 100 & 100 & 100 & 100 & 100 & 100 & 100 & 100 \\
                        & 100 & 5  & 0.31 & 100 & 100 & 100 & 100 & 100 & 100 & 100 & 100 & 100 & 97  & 100 & 100 \\
                        & 100 & 0  & 0.02 & 9   & 9   & 3   & 9   & 4   & 4   & 2   & 4   & 1   & 0   & 2   & 0 \\
                        & 200 & 50 & 1    & 100 & 100 & 100 & 100 & 100 & 100 & 100 & 100 & 100 & 100 & 100 & 100 \\
                        & 200 & 15 & 1    & 100 & 100 & 100 & 100 & 100 & 100 & 100 & 100 & 100 & 100 & 100 & 100 \\
                        & 200 & 10 & 0.88 & 100 & 100 & 100 & 100 & 100 & 100 & 100 & 100 & 100 & 100 & 100 & 100 \\
                        & 200 & 5  & 0.29 & 100 & 100 & 100 & 100 & 100 & 100 & 100 & 100 & 100 & 100 & 100 & 100 \\
                        & 200 & 0  & 0.01 & 8   & 10  & 5   & 8   & 4   & 5   & 2   & 6   & 1   & 1   & 0   & 2 \\
\hline
$X_1 \sim N(0,1)$       & 20  & 50 & 1    & 100 & 100 & 100 & 100 & 100 & 100 & 100 & 100 & 100 & 100 & 100 & 100 \\
$X_2 \sim N(0,1)$       & 20  & 15 & 0.95 & 100 & 100 & 100 & 100 & 100 & 99  & 100 & 100 & 99  & 97  & 100 & 100 \\
                        & 20  & 10 & 0.83 & 97  & 95  & 100 & 100 & 96  & 90  & 99  & 99  & 88  & 78  & 99  & 97 \\
                        & 20  & 5  & 0.4  & 55  & 48  & 94  & 90  & 46  & 41  & 88  & 72  & 31  & 19  & 69  & 54 \\
                        & 20  & 0  & 0.14 & 14  & 10  & 10  & 6   & 4   & 2   & 5   & 3   & 1   & 0   & 1   & 0 \\
                        & 50  & 50 & 1    & 100 & 100 & 100 & 100 & 100 & 100 & 100 & 100 & 100 & 100 & 100 & 100 \\
                        & 50  & 15 & 0.99 & 100 & 100 & 100 & 100 & 100 & 100 & 100 & 100 & 100 & 100 & 100 & 100 \\
                        & 50  & 10 & 0.9  & 100 & 100 & 100 & 100 & 100 & 100 & 100 & 100 & 100 & 100 & 100 & 100 \\
                        & 50  & 5  & 0.34 & 97  & 89  & 99  & 100 & 94  & 84  & 99  & 99  & 81  & 67  & 99  & 99 \\
                        & 50  & 0  & 0.05 & 14  & 12  & 13  & 13  & 6   & 7   & 6   & 7   & 1   & 2   & 0   & 1 \\
                        & 100 & 50 & 1    & 100 & 100 & 100 & 100 & 100 & 100 & 100 & 100 & 100 & 100 & 100 & 100 \\
                        & 100 & 15 & 1    & 100 & 100 & 100 & 100 & 100 & 100 & 100 & 100 & 100 & 100 & 100 & 100 \\
                        & 100 & 10 & 0.89 & 100 & 100 & 100 & 100 & 100 & 100 & 100 & 100 & 100 & 100 & 100 & 100 \\
                        & 100 & 5  & 0.31 & 100 & 100 & 100 & 100 & 100 & 100 & 100 & 100 & 100 & 97  & 100 & 100 \\
                        & 100 & 0  & 0.02 & 9   & 9   & 5   & 9   & 4   & 4   & 3   & 4   & 1   & 0   & 0   & 0 \\
                        & 200 & 50 & 1    & 100 & 100 & 100 & 100 & 100 & 100 & 100 & 100 & 100 & 100 & 100 & 100 \\
                        & 200 & 15 & 1    & 100 & 100 & 100 & 100 & 100 & 100 & 100 & 100 & 100 & 100 & 100 & 100 \\
                        & 200 & 10 & 0.88 & 100 & 100 & 100 & 100 & 100 & 100 & 100 & 100 & 100 & 100 & 100 & 100 \\
                        & 200 & 5  & 0.29 & 100 & 100 & 100 & 100 & 100 & 100 & 100 & 100 & 100 & 100 & 100 & 100 \\
                        & 200 & 0  & 0.01 & 8   & 10  & 9   & 8   & 4   & 5   & 6   & 6   & 1   & 1   & 1   & 2 \\
\hline
$X_1 \sim Cauchy(0,1)$  & 20  & 50 & 1    & 100 & 100 & 95 & 100  & 100 & 100 & 92  & 100 & 100 & 100 & 91  & 100 \\
$X_2 \sim Cauchy(0,1)$  & 20  & 15 & 0.95 & 100 & 100 & 81 & 100  & 100 & 99  & 73  & 100 & 99  & 97  & 62  & 100 \\
                        & 20  & 10 & 0.83 & 97  & 95  & 71 & 100  & 96  & 90  & 64  & 99  & 88  & 78  & 47  & 97 \\
                        & 20  & 5  & 0.4  & 55  & 48  & 42 & 90   & 46  & 41  & 33  & 72  & 31  & 19  & 19  & 54 \\
                        & 20  & 0  & 0.14 & 14  & 10  & 12 & 6    & 4   & 2   & 7   & 3   & 1   & 0   & 2   & 0 \\
                        & 50  & 50 & 1    & 100 & 100 & 98 & 100  & 100 & 100 & 95  & 100 & 100 & 100 & 93  & 100 \\
                        & 50  & 15 & 0.99 & 100 & 100 & 81 & 100  & 100 & 100 & 72  & 100 & 100 & 100 & 62  & 100 \\
                        & 50  & 10 & 0.9  & 100 & 100 & 66 & 100  & 100 & 100 & 63  & 100 & 100 & 100 & 52  & 100 \\
                        & 50  & 5  & 0.34 & 97  & 89  & 43 & 100  & 94  & 84  & 38  & 99  & 81  & 67  & 25  & 99 \\
                        & 50  & 0  & 0.05 & 14  & 12  & 12 & 13   & 6   & 7   & 9   & 7   & 1   & 2   & 4   & 1 \\
                        & 100 & 50 & 1    & 100 & 100 & 92 & 100  & 100 & 100 & 89  & 100 & 100 & 100 & 81  & 100 \\
                        & 100 & 15 & 1    & 100 & 100 & 70 & 100  & 100 & 100 & 65  & 100 & 100 & 100 & 49  & 100 \\
                        & 100 & 10 & 0.89 & 100 & 100 & 64 & 100  & 100 & 100 & 54  & 100 & 100 & 100 & 35  & 100 \\
                        & 100 & 5  & 0.31 & 100 & 100 & 32 & 100  & 100 & 100 & 25  & 100 & 100 & 97  & 20  & 100 \\
                        & 100 & 0  & 0.02 & 9   & 9   & 7  & 9    & 4   & 4   & 5   & 4   & 1   & 0   & 5   & 0 \\
                        & 200 & 50 & 1    & 100 & 100 & 86 & 100  & 100 & 100 & 82  & 100 & 100 & 100 & 77  & 100 \\
                        & 200 & 15 & 1    & 100 & 100 & 62 & 100  & 100 & 100 & 55  & 100 & 100 & 100 & 43  & 100 \\
                        & 200 & 10 & 0.88 & 100 & 100 & 49 & 100  & 100 & 100 & 42  & 100 & 100 & 100 & 35  & 100 \\
                        & 200 & 5  & 0.29 & 100 & 100 & 30 & 100  & 100 & 100 & 26  & 100 & 100 & 100 & 14  & 100 \\
                        & 200 & 0  & 0.01 & 8   & 10  & 5  & 9    & 4   & 5   & 4   & 7   & 1   & 1   & 2   & 2 \\
\hline
\end{tabular}}
\renewcommand{\baselinestretch}{1}
\caption{Frank copula linear-linear power study: The powers of the proposed test implemented by using the Rayleigh (ART) and Pycke (APT) circular uniformity tests, the Wilks
(WT) and the empirical copula (ECT) tests are compared when simulating 100 times samples of sizes 20, 50, 100 and 200 from a linear-linear density function constructed
from a Frank copula and three different marginals (exponential, Gaussian and Cauchy). The Frank copula is defined with five different values of the dependence parameter
$\varphi$ (0, 5, 10, 15 and 50). The limit case with $\varphi=0$ corresponds to the null independence model.
\label{Frankcopulapower} }
\end{center}
\end{table}

\newpage

\begin{figure}[h]
\centering
\includegraphics[scale=.85, bb=54 144 558 648]{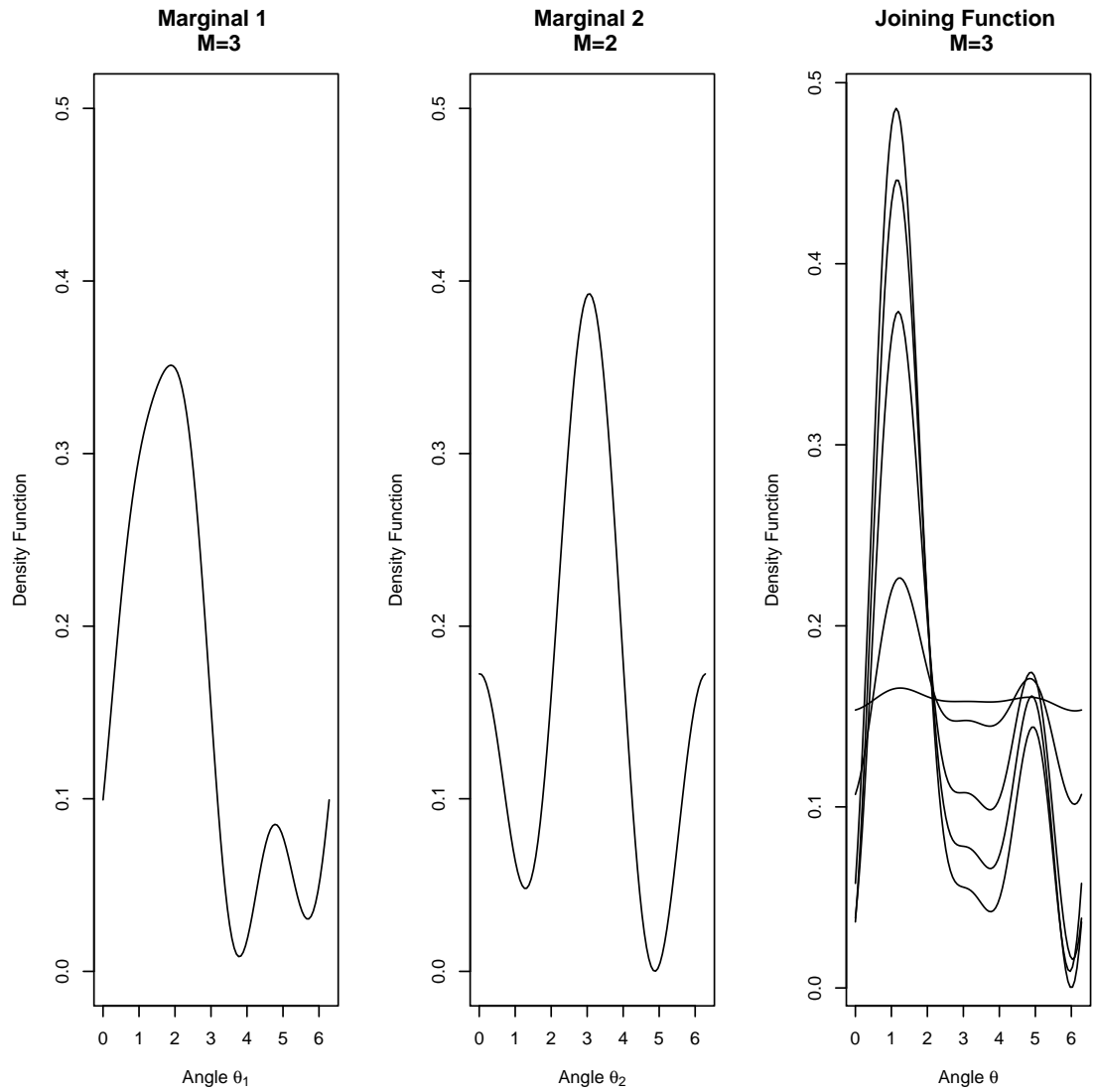}
\renewcommand{\baselinestretch}{1.00}
\caption{Circular-circular copula model: The first two plots show the marginal NNTS circular densities ($M_1=3$ and $M_2=2$) and the third plot shows the angular
(circular) joining density for different values of the parameter $c_0$ (0.7, 0.8, 0.9, 0.99 and 0.9999). The case $c_0=1$ corresponds to the circular uniform density (null
independence model).
\label{graphcircularcircularcopula}}
\end{figure}

\begin{figure}[h]
\centering
\includegraphics[scale=.85, bb=54 144 558 648]{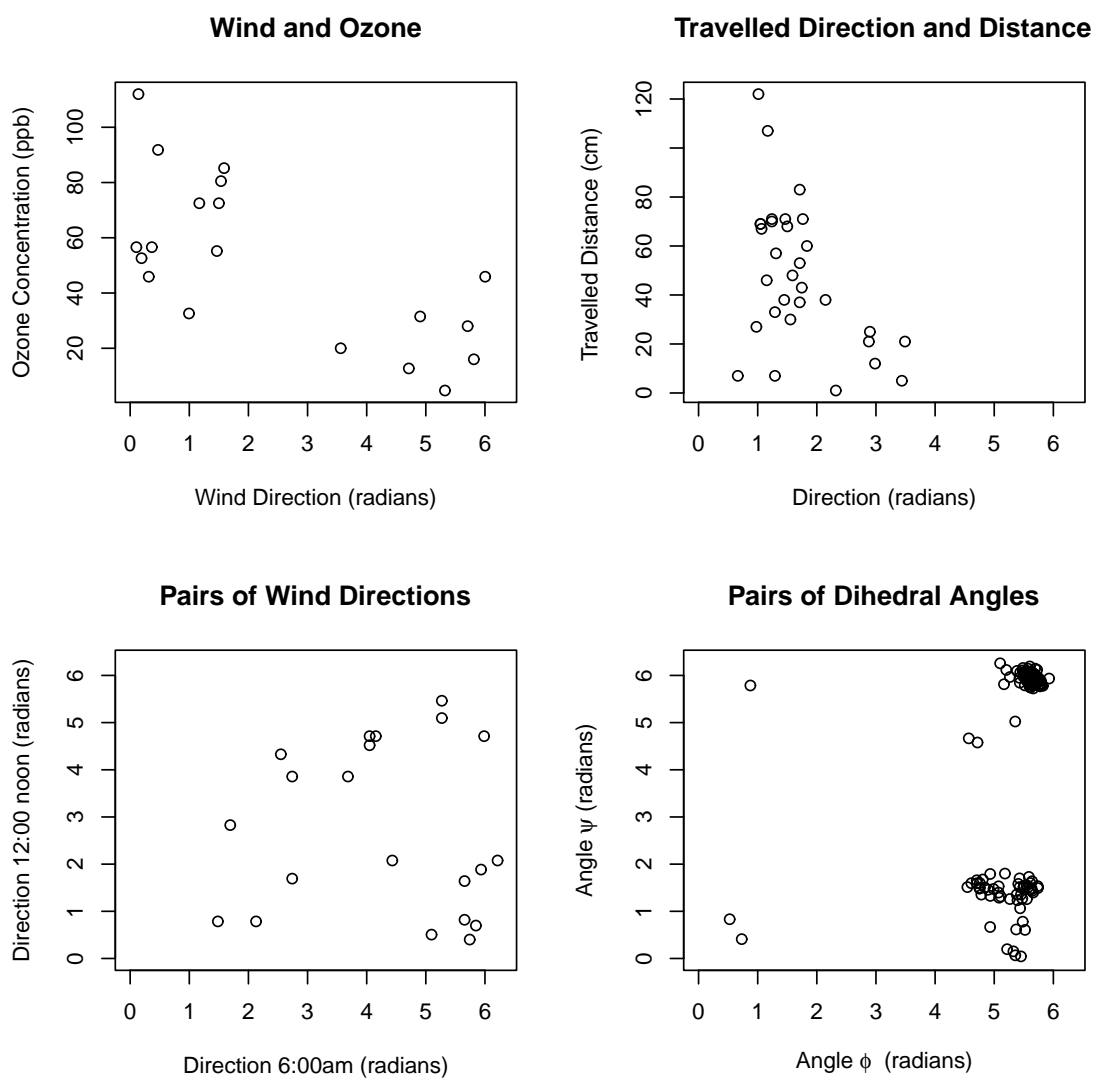}
\renewcommand{\baselinestretch}{1.00}
\caption{Scatterplots of the circular-linear (upper plots) and circular-circular (lower plots) real datasets. The upper left scatterplot shows the wind direction (relative to north) and ozone
concentration (ppb) datapoints of the dataset of Johnson and Wehrly (1977). The upper right plot corresponds to the small blue periwinkles dataset on travel distance (cm) and
direction analyzed by Fisher (1993). The bottom left plot corresponds to the Johnson and Wehrly (1977) dataset on pairs of wind directions (relative to north) at 6:00 am and 12:00 noon at a
weather monitoring station. Finally, the bottom right includes the pairs of dihedral angles in alanine-alanine-alanine segments of proteins originally analyzed by
Fern\'{a}ndez-Dur\'{a}n (2007).
\label{graphcircularcircularexamples}}
\end{figure}

\end{document}